\documentclass[a4paper, amsfonts, amssymb, amsmath, reprint, showkeys, nofootinbib, twoside]{revtex4-1}
\usepackage[english]{babel}
\usepackage[utf8]{inputenc}
\usepackage[colorinlistoftodos, color=green!40, prependcaption]{todonotes}
\usepackage[pdftex, pdftitle={Article}, pdfauthor={Author}]{hyperref} 
\begin{document}
\title{A portable atom gravimeter operating in noisy urban environments}


\author{Bin Chen$^{1, 2}$}
\author{Jinbao Long$^{1, 2}$}
\author{Hongtai Xie$^{1, 2}$}
\author{Chenyang Li$^{1, 2}$}
\author{Luokan Chen$^{1, 2}$}
\author{Bonan Jiang$^{1, 2}$}
\author{Shuai Chen$^{1, 2, *}$}
\affiliation{$^*$Corresponding author: shuai@ustc.edu.cn}
\affiliation{$^1$Hefei National Laboratory for Physical Sciences at Microscale and Department of Modern Physics,
	University of Science and Technology of China, Hefei, Anhui 230026, China}
\affiliation{$^2$Shanghai Branch, CAS Center for Excellence and Synergetic Innovation Center in Quantum Information and Quantum Physics, University of Science and Technology of China, Shanghai 201315, China}


\begin{abstract}
The gravimeter based on atom interferometry has potentially wide applications on building the gravity networks, geophysics as well as gravity assisted navigation.
Here, we demonstrate experimentally a portable atom gravimeter operating in the noisy urban environment.
Despite the influence of noisy external vibrations, our portable atom gravimeter reaches a sensitivity as good as $65\mu\textrm{Gal}/\sqrt{\textrm{Hz}}$ and a resolution of $1.1\mu\textrm{Gal}$ after $4000$ s integration time, being comparable to state-of-the-art atom gravimeters.
Our achievement paves the way for bring the portable atom gravimeter to field applications, such as gravity survey on a moving platform.
\end{abstract}


\maketitle

Atom gravimeter offers a new concept for both very sensitive and accurate absolute gravity measurement\cite{AbsoluteGrav ten meter,AbsoluteGrav Humboldt}.
This technology can reach a best short-term sensitivity of a few $\mu\textrm{Gal}/\sqrt{\textrm{Hz}}$ that outperforms state-of-the-art classical corner cube sensors\cite{AbsoluteGrav Wang,AbsoluteGrav Tino0,AbsoluteGrav Zhan0,AbsoluteGrav Landragin,AbsoluteGrav Charriere,AbsoluteGrav Tino,AbsoluteGrav Zhan,AbsoluteGrav Altin,AbsoluteGrav Bidel,AbsoluteGrav Zhanm,AbsoluteGrav Peters,zhou2013,AbsoluteGrav Lin,AbsoluteGrav Lin2}.
Nowadays, the development of portable atom gravimeters overcomes many limitations of existing sensors and open wide area of potential applications in geophysics, resource finding, gravity field mapping, Earthquake prediction as well as gravity-aided navigations~\cite{volcanic1,volcanic2,earthquake1,earthquake2,water storage1,water storage2,exploring,ice melting,navigation1,navigation2}.
More and more trials efforts have been made in the field applications of atom gravimeters recently.
Muquans has deployed a commercial portable atom gravimeter on Mount Etna for monitoring the gravity anomaly induced by the volcanic activity~\cite{Etna}.
Mounted on an inertial stabilized platform, an atom gravimeter was applied to measured the gravity in Atlantic as a marine gravimeter~\cite{ship}.
A vehicle atom gravimeter is also operated for the gravity survey in the hills~\cite{vehicle}.

Despite the high performance obtained in very quiet and well controlled laboratory conditions, such as cave laboratory\cite{zhou2013}, remote locations\cite{AbsoluteGrav Humboldt}, and underground galleries\cite{deserted,Underground Laboratory}, rare precise measurements with portable atom gravimeters are reported in the noisy urban environment with high sensitivity and resolution, since the performance is largely limited by parasitic vibrations from the ground\cite{AbsoluteGrav Bidel,AbsoluteGrav Zhanm,AbsoluteGrav Lin}.

Rising to the challenge of operating a portable atom gravimeter in the noisy urban environment, we develop a miniatured atom sensor mounted on a mobile active vibration isolation stage.
The portable atom gravimeter is then transported to and operating in the urban environment for more than 10 days.
With the external vibrations of the Raman mirror being suppressed simultaneously in three dimensions,
the vibration noise is reduced by a factor up to 2000 from 0.01 Hz to 10 Hz in vertical direction and by a factor up to 30 horizontally,
allowing our portable atom gravimeter to reach a sensitivity of $65\mu\textrm{Gal}/\sqrt{\textrm{Hz}}$ ($74\mu\textrm{Gal}/\sqrt{\textrm{Hz}}$) at night (daytime) and a resolution of  $1.1\mu$Gal after 4000 seconds integration.

\begin{figure*}[htb]
	\begin{center}
		\includegraphics[width=0.85\linewidth]{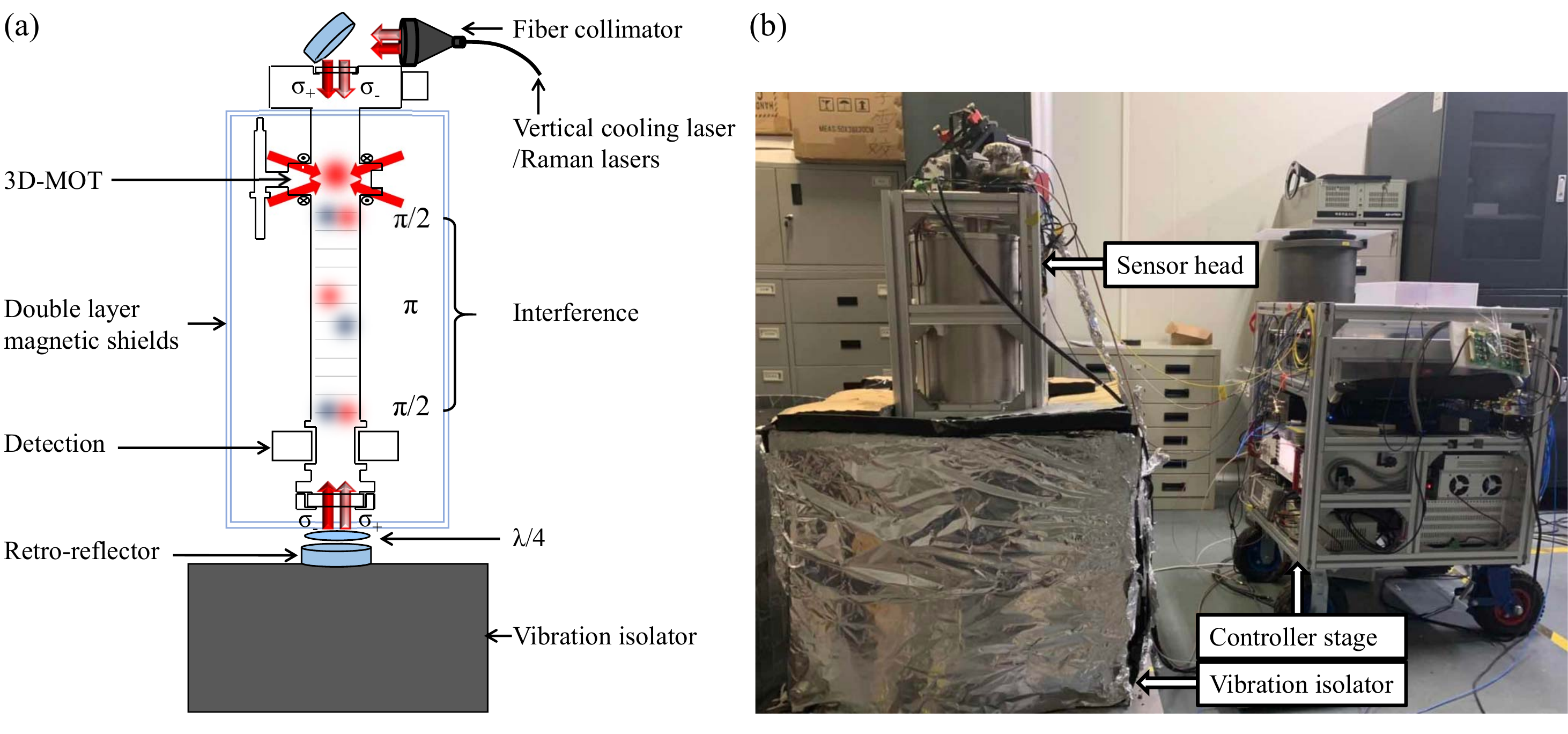}
		\vspace{-0.5cm}
		\caption{\label{fig_Setup}\text{(a)} Schematic diagram of the main science package, where the miniatured atom sensor is mounted on the portable active vibration isolation platform. \text{(b)} Photo of the portable atom gravimeter running in a noisy lab.
		}
	\end{center}
\end{figure*}

The scheme of the portable atom gravimeter operating in the urban environment is shown in Fig.~\ref{fig_Setup}\textrm{(a)}.
It includes a cold atom sensor head (30cm$\times$30cm$\times$65cm) that providing the gravity measurement and a mobile active vibration isolator (60cm$\times$60cm$\times$50cm) that suppressing the parasitic vibration from the ground.
An integrated controller package (56cm$\times$68cm$\times$72cm) provides all the lasers for manipulating cold atoms and performs data acquisition and processing.
The total power consumption is less than 400W.

The miniatured atom sensor consists of a titanium vacuum chamber, magnetic coils, optics for delivering laser beams and collectors of fluorescence signals.
The sensor is implemented in a compact 2-layer magnetic-shield, with residue magnetic field below 50 nT near the center.
The $^{87}$Rb atoms as test mass are loaded directly from the background vapor by the 3-dimensional magneto-optical trap (3D MOT) in 120ms and further cooled down to 3.7 $\mu$k by optical molasses.
Two optical-phase-locked Raman lasers are combined together and aligned carefully along vertical direction with a retro-reflected configuration (Fig.~\ref{fig_Setup}\textrm{(a)}).
The initial state is prepared by a series of Raman $\pi$-pulses, during which 10$^{6}$ atoms are selected with temperature of 300 nK in vertical direction and prepared in the magnetic insensitive state $|F=1,m_{F}=0\rangle$.
The Mach-Zehnder type matter wave interferometry is realized by shining a $\pi$/2-$\pi$-$\pi$/2 Raman pulse sequence with time interval of $T=82$ ms and $\pi$-pulse length $\tau$ = 20 $\mu$s. To prevent for the incoherent photon scattering, the Raman lasers are detuned to the red of 700MHz from the D$_2$ transition lines.
Doppler effect during the free fall is compensated by chirping the relative frequency of the Raman lasers at the rate of $ \alpha \sim $  25.1 MHz/s.
After the interferometry, the atoms at $|1,0\rangle$ \& $|2,0\rangle$ are counted by collecting the fluorescence of each state, respectively.
The interferometry fringe is obtained by scanning the chirping rate $\alpha$ and the gravity value $g$ is achieved via full-fringe fitting.

\begin{figure*}[htb]
	\begin{center}
		\includegraphics[width=0.8\linewidth]{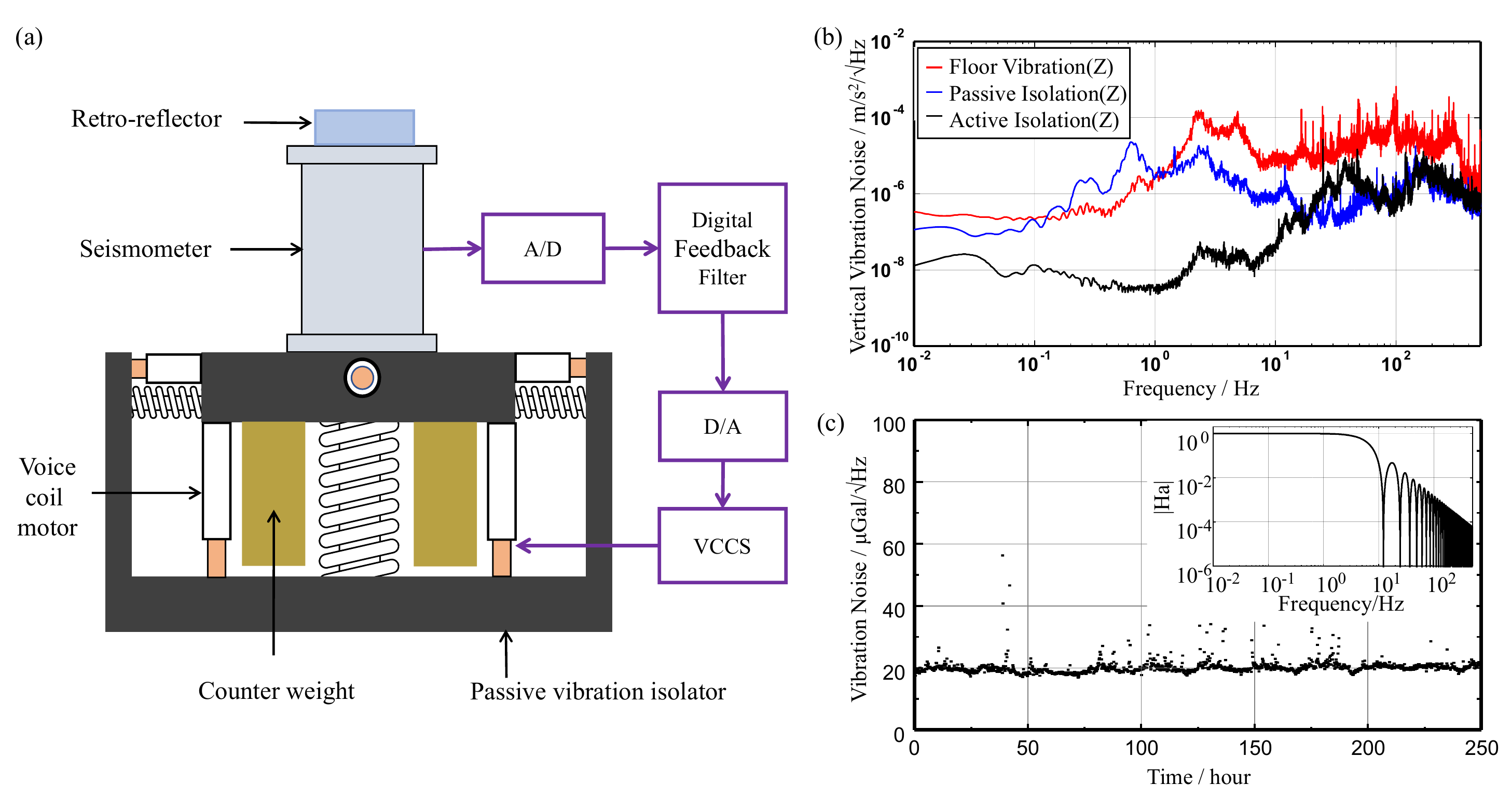}
		\vspace{-0.5cm}
		\caption{\label{fig_ThreeVIS}
			(a) Schematic overview of the three-dimension active vibration isolator.
			(b) Vibration noise on vertical direction. Red: The noise spectrum measured directly on the lab floor. Blue: The residual vibration noise on the passive isolator. Black: The residual vibration noise on the isolator with active feedback.
			(c) Long-term performance of the isolator. Inset: vibrational transfer function of the atom sensor.
		}
	\end{center}
\end{figure*}

A homemade portable three-dimension active vibration isolator underneath the atom sensor is applied to isolate the Raman retro-reflector from the ground vibration.
The schematic overview of the active vibration isolator is shown in Fig.~\ref{fig_ThreeVIS} \text{(a)}.
It is based on a commercial passive vibration isolation platform (Minus-K Technology 50BM-4). Eight voice coil motors driven by homemade voltage controlled current sources (VCCS) are added inside to implement the active feedback units. A precision three-axis seismometer (Guralp CMG-3ESP) mounted on the isolation platform is used to monitor the vibration noise of the retro-reflector. The residual vibration noises go through an anolog-digital converter are transferred to a programmable digital feedback filter with a sampling rate of 1 kHz. By passing through a digital-analog converter and the VCCS, the output signal from the filter is turned into current to drive the voice coils that implement the feedback force in three dimensions~\cite{Artical}.

The performance of the active vibration isolator is characterized by monitoring the residual vibrations with the seismometer.
Fig.~\ref{fig_ThreeVIS}\text{(b)} show the equivalent vertical acceleration noise power spectrum in different situations. 
The red curve indicates the ground vibration noise at the assembly site, located in one of the text labs at Shanghai institute, University of Science and Technology of China.
As the site is suffered badly from heavy human activities and highway traffic nearby, the vibration noise below 10 Hz is much larger than 2$\times$10$^{-7}$ m/s$^{2}$$/\sqrt{\rm Hz}$, and the vibration eigenmode around 2.5 Hz \cite{Freier} is as large as 10$^{-4}$ m/s$^{2}$$/\sqrt{\rm Hz}$.
Transformed to the noise of the gravity measurement, it is equivalent to the noise of higher than $15000\mu\text{Gal}/\sqrt{\text{Hz}}$~\cite{cheinet2008}, which is impossible for the precision $g$ measurement at $\mu\text{Gal}$ level.
It means we are in a quite noisy urban environment compared to a well established gravity measurement sites~\cite{AbsoluteGrav Peters,zhou2012,gouet2008,merlot2009}.
The blue curve indicates the residual vibration noise on the passive vibration isolation platform without feedback. The noise above 1 Hz is effectively suppressed.
While, around the natural resonance frequency of the isolation platform, 0.5Hz, the noise is slightly amplified.
The black curve indicates the residual vibration noise of the Raman retro-reflector with the active feedback loop turned on.
The residual vibration noise is reduced by a factor of 2000 from 0.01 Hz to 10 Hz, as the residual noise to crosstalk between horizontal and vertical directions is further suppressed by extra horizontal feedback channels. The vibration noise is also reduced horizontally by a factor of 30 from 0.01 Hz to 10 Hz (not shown here)~\cite{Artical}. The long-term performance of the three-dimension active vibration isolator is presented in Fig.~\ref{fig_ThreeVIS} \text{(c)}.
We characterize it by monitoring the sensitivity suffered to the residual vibration noise for more than 10 days. It is maintained around 20 $\mu$Gal$/\sqrt{\rm Hz}$ with a standard deviation of 5 $\mu$Gal$/\sqrt{\rm Hz}$.
There are some outliers (mostly during the daytime) due to occasional human activities close to the setup.
After that, the active vibration isolator recovers very quickly and works well for the whole period.

\begin{figure}[htb]
	\begin{center}
		\includegraphics[width=\linewidth]{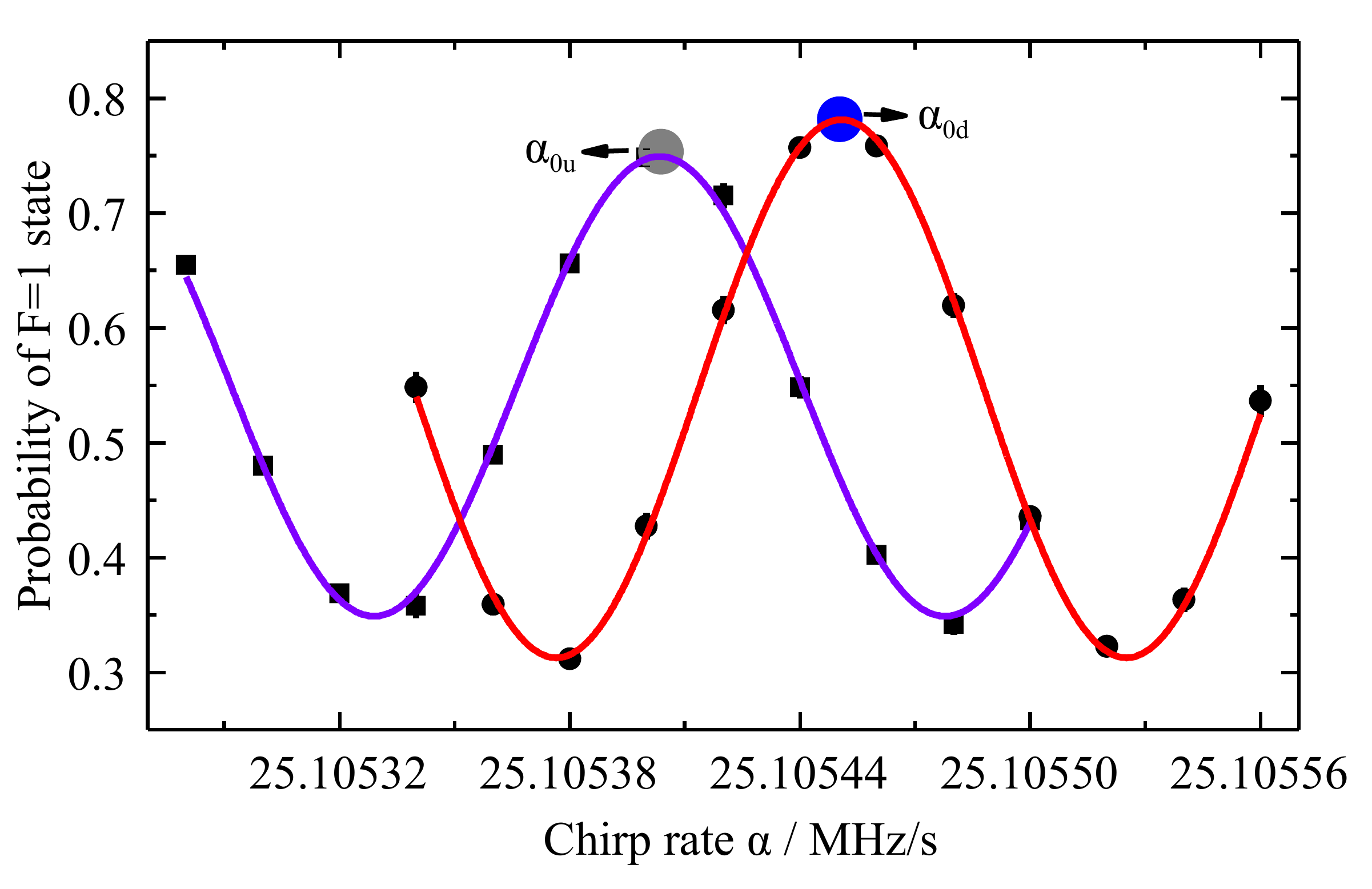}
		\vspace{-0.5cm}
		\caption{\label{fig_interference}Interferometry fringe for $T=82$ ms. It is obtained by 48 drops in 16 s for chirp up and down, respectively. Each black dot is the probability of atoms in $|1,0\rangle$ state by averaging of 4 drops. The error bar represent the statistical error. The purple and red line are the fitting according to chirp up and down, respectively.
		}
	\end{center}
\end{figure}

\begin{figure*}[htb]
	\begin{center}
		\includegraphics[width=0.75\linewidth]{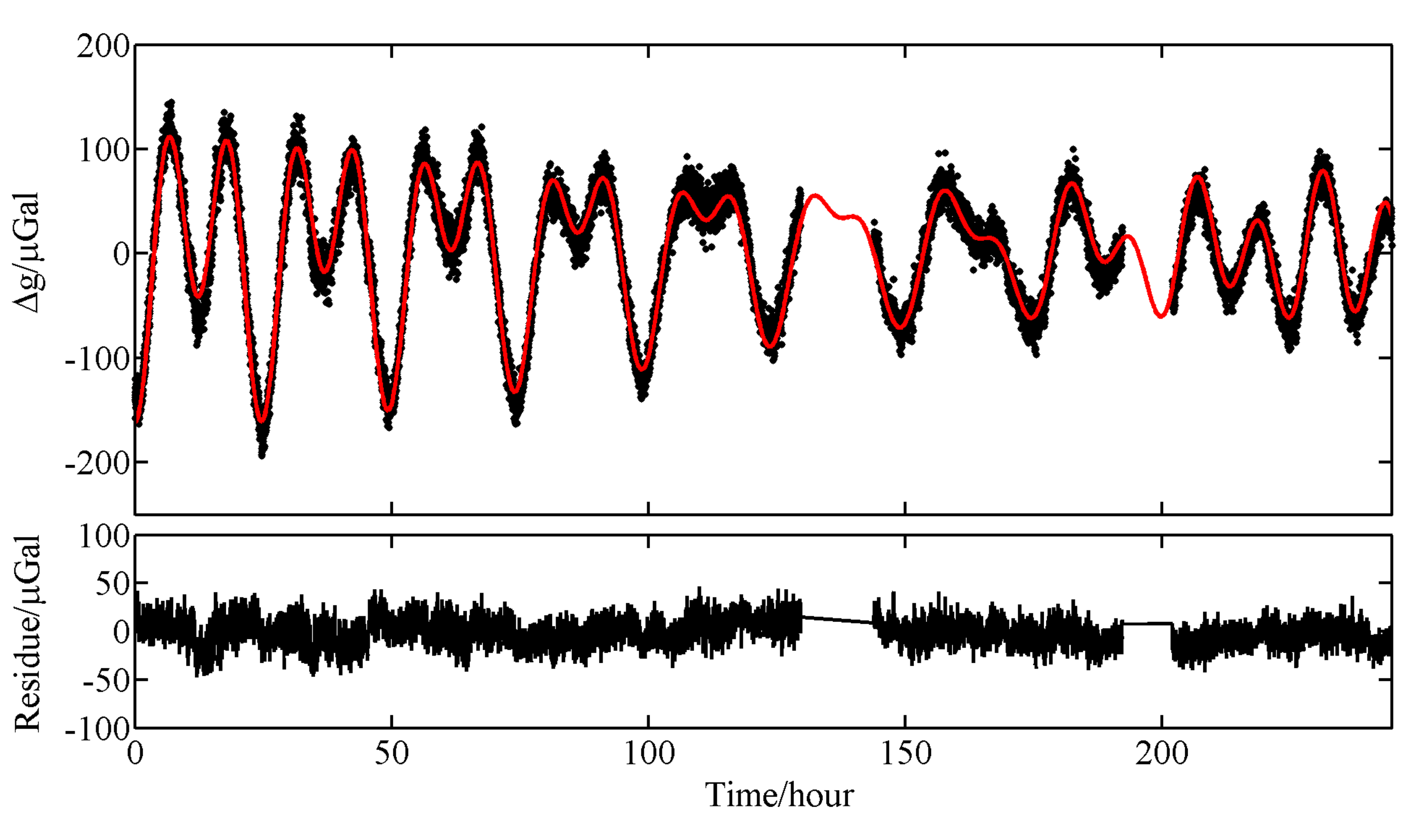}
		\vspace{-0.5cm}
		\caption{\label{fig_tide}
			top: The gravity acceleration $g$ measured by the portable atom gravimeter between 29th October and 8th November 2019. The setup works continuously for more than 10 days in the noisy lab. The two breaks (from 130 h to 143 h and from 192 h to 201 h) are cause by the lasers out of lock.
			bottom: The residue is achieved form the corresponding gravity signal substracted by Earth's tides.		 
		}
	\end{center}
\end{figure*}

\begin{figure}[htb]
	\begin{center}
		\includegraphics[width=0.9\linewidth]{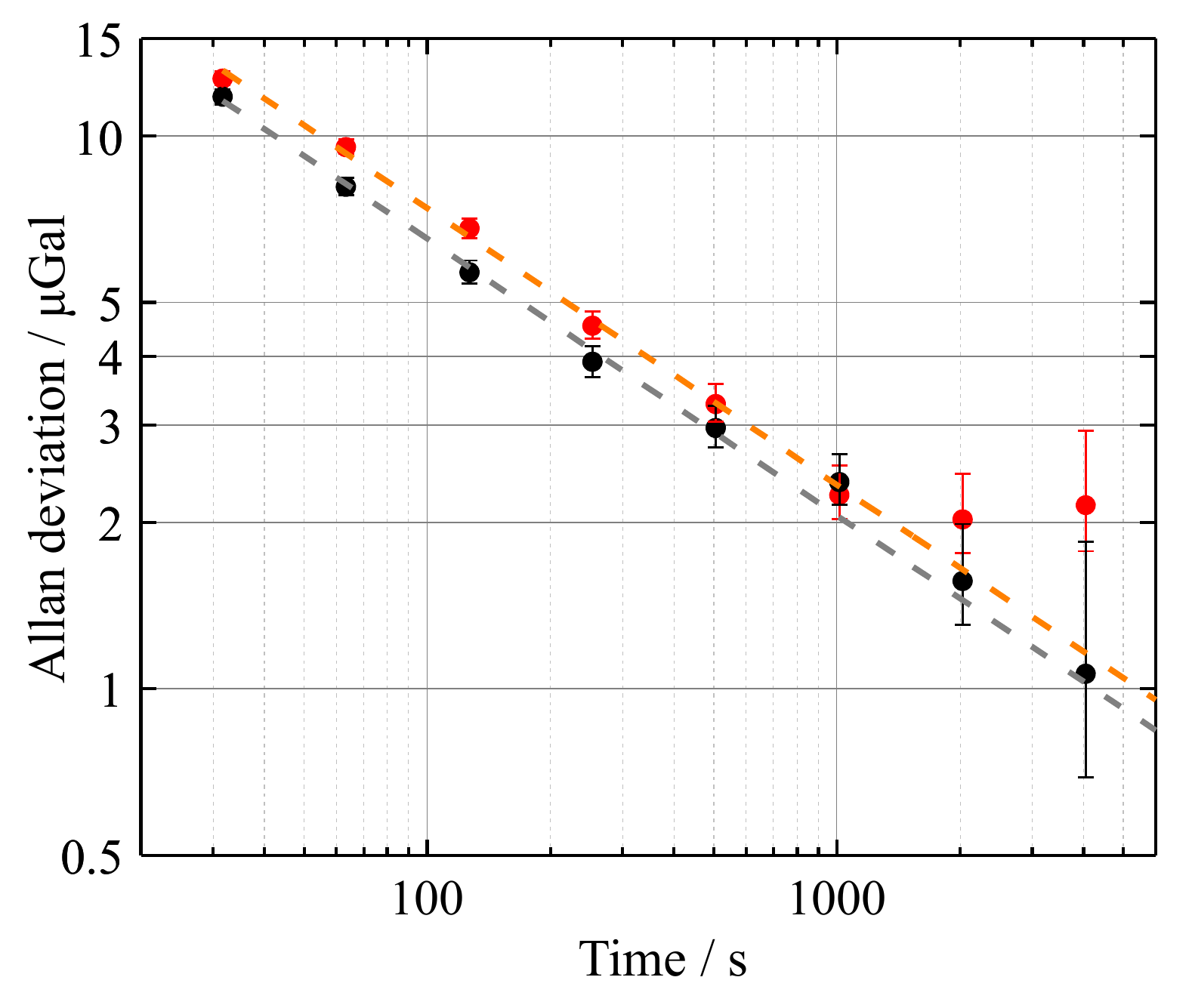}
		\vspace{-0.5cm}
		\caption{\label{fig_Allan}Allan deviation of the gravity signal corrected for earth's tides, in the daytime (red) and at night (black). The $ \tau^{-1/2} $  slopes represent the corresponding averaging expected for white noise.
		}
	\end{center}
\end{figure}

From the capture of cold atoms to the end of the interferometry, The full circle of the gravimeter takes about 300ms and works at the repetition rate of 3 Hz.
We perform the gravity measurements by $\pi/2 - \pi - \pi/2$ Raman pulses interact with the cold atoms during free-falling.
The interferometry fringe is obtained by scanning the chirping rate $\alpha$. We have
\begin{equation}
P=A(1\pm\cos(\vec {k}_{\text{eff}}\cdot \vec{g}- 2\pi\alpha)T^{2}),
\label{messi}
\end{equation}
where $P$ is normalized population of atoms in $|1,0\rangle$ state, $A$ is the contrast of the interference fringe, $\vec{k}_{\text{eff}}$ is the effective wave vector of the Raman lasers, and $T$ is the interrogation time between Raman pulses.
The gravity $g$ can be achieve via full-fringe fitting.

In order to get rid of the $\vec{k}_{\text{eff}}$-independent systematic errors, which including quadratic Zeeman shift, the one photon light shift and the radio-frequency phase shift~\cite{systematic}, we flip the direction of $\vec{k}_{\text{eff}}$ by switching the sign of chirping rate $\alpha$ for every 48 drops (16 seconds).

The interferometry fringes with the chirp up and chirp down during gravity measurements are shown in Fig.~\ref{fig_interference}. The cosine function is applied to fit the interference pattern. At each fitting extreme maximum point for chirp up and chirp down, we obtain $\alpha _ { \text {0u}}=25.105403743$MHz/s and  $\alpha _ { \text {0d}}=25.105450345$MHz/s, respectively, which mark the point with zero phase shift of the interferometer.
By the least-squares fitting of the pattern, we obtain the uncertainties of  $ \alpha _ { \text {0u} } $  and $ \alpha _ { \text {0d} } $ as $\sigma_{\alpha_{\text{0u}}}=2.9\times10^{-7}$ MHz/s and $\sigma_{\alpha_{\text{0d}}}=2.3\times10^{-7}$ MHz/s for a single interference fringe in every 16 seconds.

For obtaining $g$ value, we derive $ \alpha _ { \text {0u} } $  and $ \alpha _ { \text {0d} } $ in Eq.(\ref{messi}) with counting the systematic errors as,
\begin{equation}
\begin{matrix}
\left| \vec { k } _ { \text {eff } } \right| g T ^ { 2 } - 2 \pi  \alpha _ { \text {0u} }  T ^ { 2 } + \Delta \Phi _ { \text {dep} } + \Delta \Phi _ { \text {ind} } = 0\\
- \left| \vec { k } _ { \text {eff } } \right| g T ^ { 2 } - 2 \pi (-\alpha _ { \text {0d} })  T ^ { 2 } - \Delta \Phi _ { \text {dep} } + \Delta \Phi _ { \text {ind} } = 0
\end{matrix}
\end{equation}
where $\Delta \Phi_{ \text {dep} ( \text {ind} )  }$ represent the systematic phase shift which dependent (independent) on the direction of the $\vec k_{\textrm {eff}}$.
We derive the expression of the $g$ from those two equation and obtain:
\begin{equation}
g = \frac{ \pi (\alpha _ { \text {0u} }  +\alpha _ { \text {0d} } )}{ \left| \vec { k } _ { \text {eff } } \right|}   - \frac{\Delta \Phi _ { \text {dep} }}{  \left| \vec { k } _ { \text {eff } } \right| T ^ { 2 } }  .
\label{eq:g}
\end{equation}

The systematic phase shifts which are independent on the sign of $\vec k_{\textrm {eff}}$ are eliminated, only the $\vec k_{\textrm {eff}}$ dependent term $ \Delta \Phi _ { \text {dep} } $ is left, which includes the effect of two-photon light shift, self gravity, Coriolis forces and the wave-front aberrations et al.~\cite{systematic}.

From the Eq.~(\ref{eq:g}), we further obtain the uncertainty of each $g$ measurement:
\begin{equation*}
	\sigma _g = \frac{ \pi }{ \left| \vec { k } _ { \text {eff } } \right|}  \sqrt{\sigma _ {\alpha _ { \text {0u} } } ^2 + \sigma _ {\alpha _ { \text {0d} } }^2} \approx7.2\mu\text{Gal}
	\label{eq:uncertainty}
\end{equation*}
within 30 seconds integration time.

After the completion of the assembly and adjustment, the portable atom gravimeter performs the continuous measurement of the local gravity over 245 hours at the assembly site, from 29th October to 8th November 2019.
As shown in Fig.~\ref{fig_tide}, despite the influence of noisy external vibrations, the experimental results (black dots) agree well with earth's tides (red) predicted theoretically with an inelastic non-hydrostatic Earth model~\cite{tide}. The fluctuation of the residue signal may come from the temperature variation in the lab.

The Allan deviation of the residue signal is then calculated to characterize the sensitivity and long term stability of our portable atom gravimeter.
As shown in Fig.~\ref{fig_Allan},  the sensitivity of the portable gravimeter follows 74 $\mu$Gal$/\sqrt{\rm Hz}$ for up to 1000s during daytime and follows 65 $\mu$Gal$/\sqrt{\rm Hz}$ for up to 4000s during nighttime.
10 $\mu$Gal level is obtained after about 60 s of measurement and the Allan deviation continues to decrease down to 1.1 $\mu$Gal (2 $\mu$Gal ) within an integration time of 4000 s (2000s) during nighttime (daytime).

For nearby locations, the portable atom gravimeter is mobile enough to be deployed and perform gravity measurements in the field with handle by only one or two people. Meanwhile, it can also fit inside a miniVan and be robust enough to be transported in the long range.
Even after over 1300-km-long transportation from Shanghai to the comparison site at National Institute of Metrology of China in Changping, Beijing, our gravimeter still exhibit an $2\sigma$ uncertainty of 15.6 $\mu$Gal and reach a degree of equivalence of -12.5 $\mu$Gal compare with the reference value given by the FG5-type gravimeter NIM-3A~\cite{Artical,Xiehongtai}.

In conclusion, we demonstrate that the portable atom gravimeter operates well in the noisy urban environment, with the help of a portable three-dimension active vibration isolator.
The portable atom gravimeter reaches a sensitivity as good as 65 $\mu$Gal$/\sqrt{\rm Hz}$ and a resolution of $1.1\mu$Gal within 4000 s integration.
Moreover, the setup is robust enough to be deployed in the long range, and to perform gravity measurements with handling by one or two people.
The technique demonstrated here helps us to push the portable atom gravimeter to field applications where gravity survey has to be performed in the noisy environment.
Furthermore, with the flexibility of being mounted on a vehicle or a gyro-stabled platform \cite{ship,vehicle}, the demonstration would be of interest for applications of the atom sensor using active vibration isolation to mobile gravity survey or inertial navigation.

	This work was supported by the National Key R \& D Program of China (2016YFA0301601); the National Natural Science Foundation of China (No. 11604321); and Anhui Initiative in Quantum Information Technologies (AHY120000).


\end{document}